\documentclass[twocolumn,11pt]{article}
\usepackage[a4paper,margin=1.8cm,columnsep=0.7cm]{geometry}
\usepackage{amsmath}
\usepackage{amssymb}
\usepackage{amsthm}
\usepackage{graphicx}
\usepackage{color}
\usepackage[authoryear]{natbib}

\theoremstyle{plain}

\theoremstyle{definition}

\theoremstyle{remark}

\newcommand{\affil}[1]{}
\newcommand{\received}[1]{}

\def\sep[#1]{\iffalse {\noindent {\color{blue} \tiny \texttt{#1} {\leaders\hbox{\rule{2pt}{0.4pt} } \hfill} }} \newline \indent \fi}

\makeatletter
\renewcommand\section{\@startsection{section}{1}{\z@}%
  {-3.0ex \@plus -1ex \@minus -.2ex}{1.4ex \@plus.2ex}%
  {\normalfont\large\bfseries}}
\renewcommand\subsection{\@startsection{subsection}{2}{\z@}%
  {-2.5ex\@plus -1ex \@minus -.2ex}{1.0ex \@plus.2ex}%
  {\normalfont\normalsize\bfseries}}
\renewcommand\subsubsection{\@startsection{subsubsection}{3}{\z@}%
  {-2.5ex\@plus -1ex \@minus -.2ex}{1.0ex \@plus.2ex}%
  {\normalfont\small\bfseries}}
\makeatother

\begin{document}

\twocolumn[{%
\begin{center}
  {\LARGE\bfseries Realtime price impact detection\par}
  \vspace{1.0em}
  {\large Ilija I.\ Zovko\par}
  \vspace{0.50em}
  {\small ilija.zovko@algoquants.co.uk\par}
  \vspace{0.30em}
  {\small Draft\par}
  \vspace{1.2em}
  \begin{minipage}{0.86\textwidth}
  \setlength{\parindent}{1.5em}
  \centerline{\textbf{Abstract}}
  \medskip\small

\sep[The problem]
An important question for an algo trader working an order is to understand if their actions are moving the market against them -- i.e., \emph{causing} market impact.  The conventional answer usually is one of two: (i) monitor price slippage in real-time, potentially reducing adverse activity with increased slippage, or (ii) do away with dynamic trading adjustments and rely on semi-static rules based on \emph{ex-post} estimates of slippage over a large sample of events.

Both of these approaches fail to capture the dynamic nature of “quality” of liquidity in markets. At one moment an action in the market (say a limit order post on the near side) could attract a fill from a VWAP algo trading in the opposite direction, at another moment it could trigger an adverse price move when someone pulls liquidity from the far side.

Realtime monitoring fails because reliably estimating slippage is statistically expensive -- it requires hundreds of fills before it can be told apart from background volatility. More fundamentally however, it does not establish \emph{causality}. Observed adverse price moves may be caused by the trader's own actions, or by an unrelated participant competing for the same liquidity and capturing the same alpha. The optimal response (say, slow down vs.\ speed up) is opposite in the two cases.

\sep[The method]
We propose a method that detects price impact, on a per-action basis, by measuring the \emph{timing synchronicity} between a trader's actions and subsequent adverse market events. The method at heart is a test for statistical \emph{surprise} in the timing of adverse  events post trader action.

We must be clear in that we do make a leap of faith here and \emph{assume} that surprisingly fast adverse market events are evidence of causation and that the action triggered them -- a direct signature of impact and information leakage.

We are careful to present this as a \emph{proposal} for a potentially useful metric rather than a validated one. The contribution rests on an assumed linkage -- that a surprising action-to-event timing is a signature of, and predicts, adverse price impact. The statistical machinery (calibration and detection power) can be, and is, demonstrated by simulation, but this linkage itself cannot be established by simulation. Any simulated market would simply encode the assumption. Validating it requires real execution data; we set out the empirical tests that would do so.

\sep[Details]
By treating the market events as a Poisson process with locally estimated, time-varying intensity, each trader action and subsequent market events yield a $p$-value quantifying how ``surprising'' it is to observe a market event so soon after the action.

We derive the predictive duration density from both a Bayesian and a frequentist standpoint, show they coincide, and use Fisher's method to optimally combine the $p$-values of successive actions so that reliable evidence accumulates from only a handful of actions.

The result is a statistic that is computable in real time, reacts within a few actions or fills rather than a few hundred, and -- unlike slippage -- is built around the causal question rather than mere correlation. It allows a trader to begin an execution with say a broad market access or algos and to throttle exposure to specific venues, counterparties or trading styles \emph{if and when} impact is actually detected.

\end{minipage}
\end{center}
\vspace{1.4em}
}]

\section{Introduction}
\sep[Impact is the central cost of execution]
In certain situations, the dominant cost of executing a large order is not the spread or the commission but the \emph{market impact} the order itself generates: the adverse price drift caused by the order revealing information about the trader's intent to the rest of the market. Once other participants infer that a large buyer (or seller) is active, they reposition ahead of the residual order, and the trader systematically pays worse prices on the unexecuted remainder. Controlling impact is therefore the central problem of execution.

\sep[Detection, not just minimisation]
A large body of literature addresses how to \emph{schedule} an order to minimise expected impact in advance -- in many cases, assuming market liquidity to be static. In reality, however, available liquidity is highly dynamic: market makers with different axes and inventory levels, manual traders, algos executing TWAPs or VWAPs, all enter and exit the market at various points in time. There are also short-term traders trying to infer the presence of large orders to profit from their (almost) inevitable price impact. 

The same venue or order type that was benign a minute ago may now be dominated by participants who react to the trader's actions adversely. An execution algorithm that cannot sense impact as it happens cannot react to it, and is forced to fall back on static, pre-committed schedules or conservative actions that sacrifice liquidity access or are outright unprofitable.

The terms “market impact”, “trading actions” or "events" can be thought of in broad terms. E.g., liquidity being removed from the far side as a consequence of a trader placing a limit on the near side is an example of a trading action causing market impact. For simplicity of exposition however, going forward we will simply define market impact -- the market's response to trader actions -- as “adverse price moves caused by trader fills”. 

\sep[The reactive requirement]
The operationally pressing question therefore is: \emph{while the order is live, how does the trader know whether impact is actually occurring as a consequence of their fills?} 

What is needed is a detection statistic with the following properties. First, it must be \emph{reactive} and computable realtime. It must deliver a usable signal after only a small number of fills. There is no point knowing the execution was costly only after it is done.

Second, it must relate to \emph{causality}: it should distinguish \emph{unexpected price moves the trader's own fills are causing} from \emph{adverse price moves driven by unrelated activity or alpha}. The correct trading response is opposite in the two cases. 

If the adverse price moves are caused by the fills, the appropriate response might be to slow down and reduce participation. If the adverse price moves are caused by alpha or by another market participant consuming the liquidity, the response might be to speed up trading. Very different outcomes!\footnote{The execution setting here is commonly understood as a relatively slow alpha resulting in sparse fills in relation to more frequent market prints. It does not apply to high frequency trading.}

\subsection{Joint probability of duration and impact}
\sep[The event, and what we test]
This paper sets out to propose such a statistic by starting from what a fill actually produces. When we receive a fill, the market eventually answers: the next price change arrives after some delay $\delta t$, and when it arrives it has some signed size $\delta p$. Each fill is therefore an event with two coordinates, drawn -- absent any impact -- from the market's background joint probability law
\begin{equation}
p(\delta t,\, \delta p).
\label{eq:joint}
\end{equation}
Impact distorts this law: a "toxic" fill tends to be followed by an adverse price change that comes \emph{sooner} (small $\delta t$) and moves \emph{against} us ($\delta p$). Detecting impact is detecting that the post-fill events are no longer consistent with the background.

We build the detector in two steps, and the same two steps work along either coordinate, or both at once. First, for each fill we score how surprising its event is under a \emph{locally estimated} null -- a $p$-value, the probability under the background law of an event at least as extreme. Second, a single surprising fill proves little, so we combine the per-fill $p$-values of a handful of fills by Fisher's method into one continuously updated statistic. The contribution is this two-step methodology; the choice of axis is a modelling decision within it.

The two axes are not interchangeable, and the asymmetry between them is the one between a first-passage time and a price level. The factorisation
\begin{equation}
p(\delta t,\, \delta p) = p(\delta t)\, \, p(\delta p \mid \delta t)
\label{eq:factor}
\end{equation}
is the natural marked-point-process decomposition -- the law of \emph{when} the next change arrives, times the law of \emph{how large} it is given that timing. 

But the two operational tests are better seen as dual slices of the same event. Fixing a price move and asking \emph{when} it happens gives the duration $\delta t$ to the next change -- a first-passage time, and the object this paper tests. Fixing instead a horizon $\tau$ and asking \emph{how far} price has moved by then gives the markout, the conventional price-domain measure.\footnote{A return is always quoted at a horizon, so there is no horizon-free price marginal to test: $p(\delta p)$ on its own is merely the size law of the next change, not a return.}

The two slices answer different questions. The time test asks whether market activity is unusually synchronised with our fills. Its null is ``the market would have moved when it did regardless of us,'' a rejection points to a \emph{causal} echo. 

The markout asks only how adverse the move was, and on its own it is correlational -- the move may be caused by us, or by someone else. They are sharpest together: a change that is both \emph{fast} and \emph{adverse} is the clearest signature of caused impact. Loosely, time supplies causality and price supplies magnitude.

At its core, then, the method is a test for \emph{surprise} -- a departure of the post-fill adverse events from their background law -- made decisive by pooling across fills.

In this paper we take the time axis. The key assumption is that impact and information leakage leave a \emph{timing} signature: when our own fills are informative to the market, market prints tend to follow them unusually quickly.

Because impact moves the price \emph{against} us, we build the test on adverse events from the outset: henceforth a market \emph{print} means a price change that moves against the fill -- up after a buy, down after a sell -- not every price change. A surprisingly fast adverse print is then the signature of impact we are after; counting all price changes would only detect generic market activity. 

For clarity, "prints" are \emph{unexpected} adverse effects of "fills" (trader actions). An \emph{expected} reduction of liquidity on the far side as a consequence of a trader's market order - is \emph{not a "print"}.

To quantify the \emph{surprise} of an adverse print we model the market print stream as a Poisson process with a slowly varying, stochastic intensity that we treat as locally constant and estimate from recent durations, and by computing, for each fill, the probability of seeing the next market print as soon as we did.

Combining these probabilities across fills with Fisher's method yields a single, continuously updated measure of how strongly the trader's activity is being echoed by the market. 

Developing the price \emph{magnitude} -- the markout -- and the full joint test is the natural next step, which we leave to future work.

\subsection{Caveats}
\sep[A proposal, and the limits of what we claim]
We need to be honest from the outset about the status of this contribution, because it governs how the rest of the paper should be read. The paper bundles together two logically distinct claims. 

The first is \emph{statistical}: that a departure of the post-fill durations from the locally estimated background process can be detected, with a calibrated false-positive rate and useful power, from only a handful of fills.

The second is \emph{economic}: that such a timing departure is a genuine signature of -- and predicts -- adverse price impact caused by the fill. 

The first claim is a property of the estimator and we verify it by simulation. The second claim, the assumed linkage between surprise in timing and impact in price, is where the real meat of the method lies.

While potentially intuitive, ultimately it is an empirical hypothesis about how markets actually behave. It cannot be settled by simulation: any simulated market that exhibited the linkage would do so only because we built it in, making the exercise circular.

We therefore present the method as a \emph{proposal for a potentially useful metric}, establish by simulation only what simulation can legitimately establish, and touch in  Section~\ref{sec:scope} on the empirical test -- correlating the timing surprise against independently measured price moves on real fills -- that would be needed to validate, or refute, the linkage itself.

\subsection{Related work}
\label{sec:related}
\sep[Precursor]
This paper generalises an earlier proposal for the special case of dark-pool fills~\citep{zovkodark}, which introduced the “surprise” in duration measured by the $p$-value and the Fisher combination to detect information leakage from dark fills. The present work reframes that construction for market impact in general -- any market activity, dark or lit, any venue or asset class.  In an appendix we also sketch out what an extension to using Hawkes self-exciting processes in place of time-varying Poisson would look like.

\sep[Toxicity metrics]
The operational goal -- detecting adverse execution conditions while an order is live -- is shared with the literature on order-flow \emph{toxicity}. A well-known metric is VPIN, the volume-synchronised probability of informed trading~\citep{vpin}, which gauges toxicity from volume imbalance accumulated in volume-time. More recently, \citet{toxicflow} predict toxic client flow with an online Bayesian neural network. Flow toxicity is also central to the practice of electronic FX dealing, where it is managed rather than merely measured, through the ``last look'' acceptance window and internalisation of client flow~\citep{oomenlastlook, oomenaggregator, butzoomen}.

These approaches differ from ours in design goal and mechanism rather than in quality. VPIN aggregates many trades into volume buckets and measures market-wide imbalance; the toxic-flow work is a supervised predictor trained on labelled outcomes.

The proposed statistic instead reads the \emph{timing} of market prints relative to the trader's \emph{own} fills, is designed to react within a handful of fills, and is built around the question of whether those fills are \emph{triggering} market activity.

\sep[Models of the joint duration--return law]
The joint probability of prices and durations \eqref{eq:joint} is not new. Its factorisation \eqref{eq:factor} is the marked-point-process decomposition behind the high-frequency econometrics of durations and returns: the autoregressive conditional duration model~\citep{englerussell1998} for the duration, and the ultra-high-frequency GARCH of \citet{engle2000} for the return given its duration -- whose central finding, that short durations accompany higher volatility, is exactly the ``market heating up'' that compresses durations as prices move. 

The same coupling drives the subordination, or business-time picture, in which returns are a Brownian motion run on a stochastic clock set by trading activity~\citep{clark1973, anegeman2000}, and the self- and mutually-exciting (marked Hawkes) models of trades and prices~\citep{bacrymuzy}.

Our aim is different from all of these. That literature \emph{models and forecasts} the joint dynamics; we use the joint law only as a \emph{null} against which to score the surprise of each post-fill event, and we estimate it \emph{locally} -- from a few recent durations -- rather than through a fully specified dynamic model.

Locality quietly absorbs the slow heating and cooling of activity that those models represent explicitly, at the cost of not exploiting it; the explicit-coupling version (the Hawkes appendix) is a step toward the latter.

\section{Why slippage is the wrong realtime statistic}
\label{sec:slippage}
The standard way to monitor impact dynamically is to measure average post-fill slippage. For a fill at time $t$ with signed direction $\epsilon_t \in \{\text{\small sell}=-1,\text{\small buy}=+1\}$ and log mid-price $p_t$, the average impact at horizon $\tau$ is
\begin{equation}
\mathcal{R}(\tau) = E_t \left[ \epsilon_t \cdot (p_{t+\tau} - p_t ) \right],
\label{eq:slippage}
\end{equation}
where the expectation averages over fills, possibly bucketed by venue, size or order type. A positive $\mathcal{R}(\tau)$ indicates the price systematically moves against the trader following a fill. While intuitive and ubiquitous, it is ill-suited to realtime application for two distinct reasons: it is statistically slow, and it is causally ambiguous.

\subsection{Slippage is statistically slow}
The difficulty with impact estimation is in the relative magnitudes of slippage and price return variance. To reliably ascertain that $\mathcal{R}(\tau)$ is different from zero, implying the existence of systematic price movement following a fill, a t-test is commonly used, comparing the mean price slippage post fill $\mu$ to the standard deviation of unconditional price returns $\sigma$. The fact that the mean grows linearly with the number of fills $T$ and the standard deviation as a square root, imposes a lower bound on the number of data points required to detect significance as
\begin{equation}
\frac{T \mu}{\sqrt{T} \sigma} \geq 1 
\end{equation}
which, solving for $T$, becomes
\begin{equation}
\sqrt{T} = \frac{1}{\mu/\sigma} 
\label{eq:Tdependency}
\end{equation}
Interpreting this as the time required to detect a signal of slippage in a noisy market environment, the minimum time is at least equal to $(\sigma/\mu)^2$ or $1/\text{Sharpe}^2$~\citep{farmer}. The smaller the Sharpe ratio, the longer it will take to significantly detect a signal of slippage, and detecting a half as strong signal takes four times longer.

\sep[Practical example]
Complicating things further is that this low bound assumes an unrealistic 100\% trade participation rate, together with a fill at each market price change. 
Using more realistic assumptions things get worse. An order-of-magnitude calculation shows that typically hundreds of fills are required to detect slippage.

Order of magnitude for impact per fill is $\mu=0.5$bp (basis points), while price return standard deviation \emph{per return} is about $\sigma_1=3$bp. Rough estimates from market data reveal that there are say 3 price changes per print (market trade). Trading an order at a very high 20\% participation, we would expect to get a fill for each 5 prints, corresponding to $3 \cdot 5=15$ price changes. The standard deviation of market returns at this time horizon is $\sigma_{15} = \sigma_1 \sqrt{15} \sim 12 \text{bp}$. Plugging these rough numbers into the above expression for the minimum number of fills required to statistically detect slippage we get
\begin{equation}
T = \frac{1}{(\mu/\sigma_{15})^2} = \left(\frac{12}{0.5}\right)^2 \sim 570 \text{ fills.}
\end{equation}

A trader must collect of order several hundred fills before slippage can be reliably distinguished from noise -- far more than most executions ever generate, and far too slow to react to. Consequently, slippage-monitoring controllers spend most of their time responding to random price drift rather than to genuine impact. 

The number \eqref{eq:Tdependency} is a \emph{generous} lower bound: it uses a one-standard-deviation threshold, whereas a genuine significance level (e.g.\ $t\approx 2$, as we use for the timing test later) multiplies the required $T$ several-fold -- into the thousands in the example above. The comparison with the timing detector is therefore conservative against our own argument.

\subsection{Slippage is causally ambiguous}
\sep[The sign of the response is not identified]
The second, and more troubling, problem is that even a cleanly measured $\mathcal{R}(\tau) > 0$ does not tell the trader what to do. Adverse slippage following a fill is consistent with two opposite states of the world. In the first, the fill itself is informative: it leaks the trader's intent, other participants react, and the price moves away. The correct response is likely to \emph{slow down} -- reduce the fill rate, raise the minimum acceptable fill size, or avoid the offending venue. 

In the second, the adverse move is driven either by correctly predicted alpha or by an unrelated participant competing for the same liquidity; the trader's fills are incidental. Pulling back here does nothing to reduce the slippage; it either fails to capture the alpha or forfeits good prices in an escaping market -- the correct response is to \emph{speed up}.

\sep[Slippage measures correlation, impact requires causation]
Slippage is a marginal, unconditional correlation between own fills and subsequent returns. It cannot, by construction, separate these two regimes, because it never asks whether the price move was \emph{triggered by} the fill. A realtime impact statistic must instead estimate causality directly. This is what the timing-based construction of the next section is designed to do.

\section{Impact leaves a timing signature}
\label{sec:premise}
\sep[Basic premise]
We now develop the first step of the approach -- the per-fill $p$-value -- on the time axis, scoring the duration $\delta t$ to the next market print against its background law. The premise is simple. Absent any impact, the arrival of market prints is governed by the prevailing background trading rate and is statistically independent of the precise instants at which the trader receives fills. If, on the other hand, a fill conveys information, the market responds, and prints tend to cluster \emph{immediately after} the fill. Each own fill is therefore a small natural experiment: we observe the duration from the fill to the next market print and ask how surprising that duration is, given the rate of market prints around that moment.

\begin{figure*}[t]
\includegraphics[width=\textwidth]{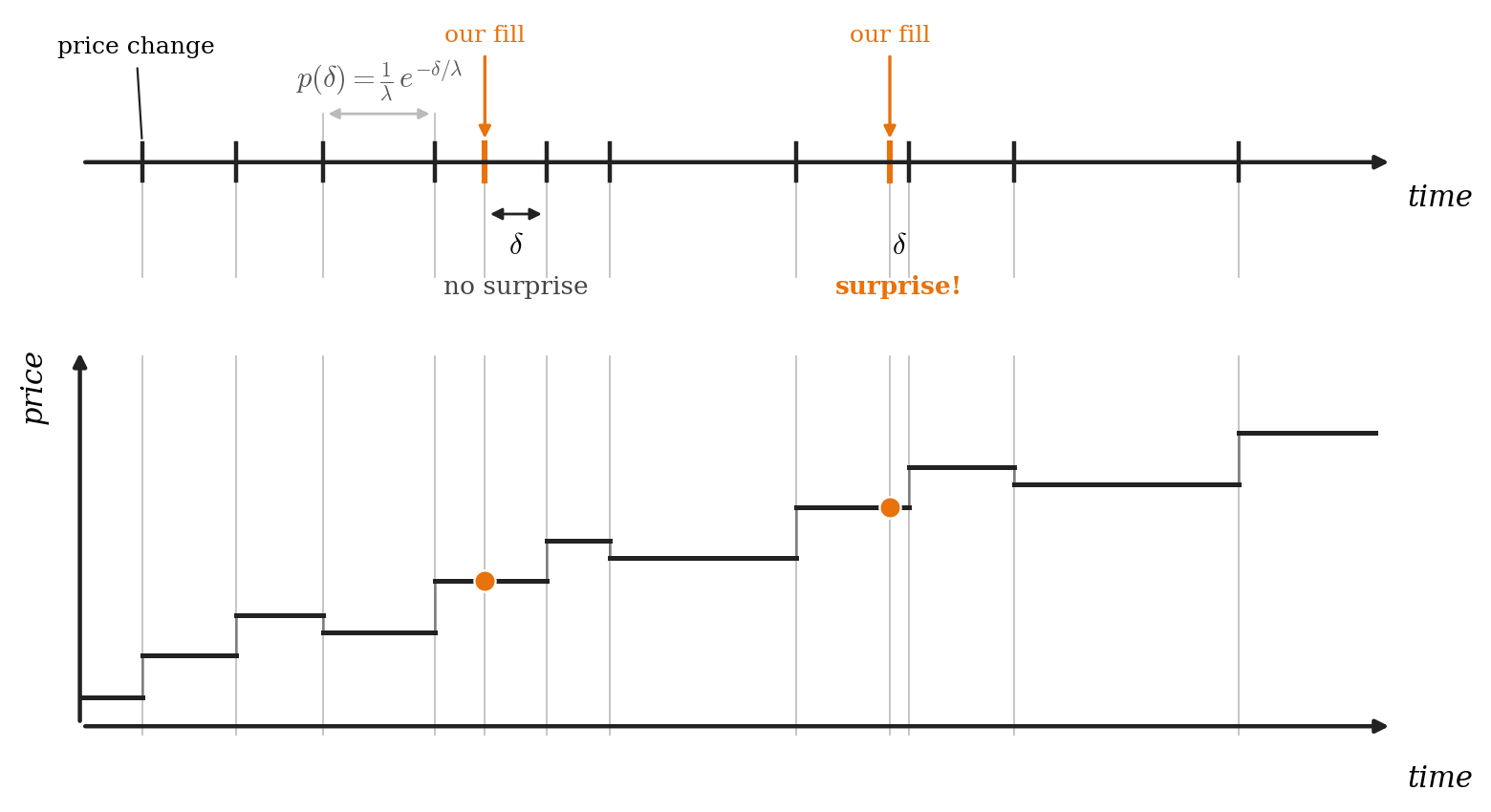}
\caption{
{\bf The timing signature of impact.}
The upper axis shows the timeline of market price-change events in time; under the no-impact null these arrive as a Poisson process whose inter-event durations are exponential, $p(\delta)=\frac{1}{\lambda}e^{-\delta/\lambda}$ with mean duration $\lambda$. Two of the trader's own fills are marked. For each fill we measure the duration $\delta$ to the next market print and judge it against the locally estimated background rate. A print that follows the fill after a typical duration (left, ``no surprise'') carries no evidence of impact; a print that follows unusually quickly given the prevailing rate (right, ``surprise'') is evidence that the fill triggered it. The lower panel shows the idealised corresponding price path, with the two fills marked; the surprising fill is the one followed by an immediate adverse move (in this illustration the fills are buys).
}
\label{fig:timedomain}
\end{figure*}

\sep[An example]
Suppose the background rate at the time of a fill is one print per second. A market print arriving $10$\,ms after our fill is then highly surprising under the background rate, and is plausibly explained as a reaction to the fill. A print arriving half a second later is unremarkable and provides no evidence of impact. The strength of the evidence in each fill is graded -- it depends continuously on how short the observed duration is relative to the local mean -- and we will quantify it as a $p$-value. Figure~\ref{fig:timedomain} illustrates the construction.

\sep[One fill is not enough]
A single surprisingly fast print proves little -- it may be coincidence. But the evidence from successive fills can be combined, and once a few fills each contribute some surprise, their joint improbability under the no-impact hypothesis becomes decisive. The number of fills needed is not fixed: the more surprising each fill, the fewer are required. Formalising this -- the per-fill $p$-value and its optimal combination -- is the content of the next sections.

\section{The per-fill surprise}
\label{sec:perfill}
\subsection{Model and notation}
\sep[The Poisson assumption]
We model market prints (excluding the trader's own fills) as an inhomogeneous Poisson point process with intensity $\lambda(t)$, slowly varying relative to the inter-print durations. Over the short window around a given fill the intensity is approximately constant, so that the durations between consecutive prints are independent and exponentially distributed,
\begin{equation}
p(\delta\mid\lambda) = \frac{1}{\lambda}\,e^{-\delta/\lambda}, \quad \delta \ge 0,
\label{eq:expdens}
\end{equation}
with mean inter-print duration $\lambda$ (equivalently print rate $1/\lambda$). The cumulative probability of a duration shorter than $d$ is
\begin{equation}
P(\delta \le d \mid \lambda) = \frac{1}{\lambda}\int_0^d e^{-\delta/\lambda}\,d\delta = 1 - e^{-d/\lambda}.
\label{eq:cumprob}
\end{equation}

A remark on the Poisson assumption is in place, since it carries a caveat of the construction. Real prints are not exactly Poisson; they cluster. It helps to separate three sources of that clustering, because the test treats them quite differently.

\emph{Exogenous rate variation} -- the rate drifting with news, time of day, or a common signal many participants act on is handled. The locally estimated $\hat\lambda$ rides the busy or quiet period, so a fast print arriving merely because the market is active is correctly \emph{not} surprising. 

\emph{Self-excitation triggered by the fill itself} -- our fill is a print, and in a responsive market a print begets more prints. This - by our definition in Section~\ref{sec:premise} - is impact.  Flagging it is a true detection, not a false alarm.

One genuine confound is the third: \emph{existing burst of activity the fill happens in}. For example, an algo placing a sell limit order during a burst of buying. The Poisson test cannot separate this from impact, because both appear as a fast print after the fill. The windowed $\hat\lambda$ is a \emph{lagging} average. The instantaneous intensity just after an event is higher. So the per-fill $p$-value \emph{over-states} significance when fills happen to fall in active periods.

The remedy is to replace $\hat\lambda$ with the Hawkes conditional intensity $\lambda(t\mid\mathcal{H}_t)$ evaluated at the fill instant. This moves exactly the part of the post-fill acceleration that the recent history already predicts into the \emph{expected} background, and leaves only the excess, which is the part most plausibly caused by the fill.  The Hawkes step is what makes the ``surprise'' estimate truly valid in a realistic market.  We nonetheless develop the Poisson case. It has useful simplifying properties and admits the closed-form solutions the Hawkes case does not. We outline the Hawkes generalisation in the appendix.

\sep[The waiting-time paradox]
A useful consequence of the memoryless property of the Poisson process is that the duration from an \emph{externally chosen} time -- such as the instant of our fill -- to the next print has the \emph{same} exponential distribution \eqref{eq:expdens} as the inter-print durations themselves. Under the null of no impact, the fill-to-next-print duration is drawn from exactly the law we estimate from the surrounding inter-print durations. This is what makes the fill-to-print duration directly comparable to the background, and is the formal basis of the test. 

\subsection{The predictive density of the next duration}
\sep[What we need]
We have observed $n$ recent inter-print durations $\delta = (\delta_1,\dots,\delta_n)$ just before the fill. Since we have \emph{not} observed the true rate $\lambda$, we must account for the uncertainty in estimating it from a short window. What we want is the predictive density of the next duration $\delta' \equiv \delta_{n+1}$ -- the fill-to-print duration -- given the observed durations, $p(\delta'\mid\delta)$. We derive it two ways.

\subsubsection{Bayesian derivation}
\sep[Likelihood, prior, posterior]
The likelihood of the observed durations under \eqref{eq:expdens} is
\begin{equation}
\mathcal{L}(\lambda;\delta) = \prod_{i=1}^{n}\frac{1}{\lambda}e^{-\delta_i/\lambda}
= \lambda^{-n}\,e^{-n\bar\delta/\lambda},
\label{eq:like}
\end{equation}
where $\bar\delta \equiv \frac{1}{n}\sum_{i=1}^n \delta_i$ is the observed mean duration. Since $\lambda$ is a scale parameter, the appropriate non-informative (Jeffreys) prior is $\pi(\lambda) \propto 1/\lambda$~\citep{jeffreys1946}.\footnote{The $1/\lambda$ prior is the standard non-informative choice for a scale parameter: it is invariant under reparameterisation (e.g.\ between the mean $\lambda$ and the rate $1/\lambda$) and under rescaling of the time unit.} The posterior is then
\begin{equation}
\pi(\lambda\mid\delta) \;\propto\; \pi(\lambda)\,\mathcal{L}(\lambda;\delta)
\;\propto\; \lambda^{-(n+1)}\,e^{-n\bar\delta/\lambda},
\label{eq:postunnorm}
\end{equation}
which is an inverse-gamma density kernel in $\lambda$.  The normalisation integral, through a change of variables $\lambda=n\bar\delta/x$, becomes
\begin{align}
\int_0^\infty \lambda^{-(n+1)}  e^{-n\bar\delta/\lambda} d\lambda =& \nonumber \\
=(n\bar\delta)^{-n}\int_0^\infty x^{n-1} e^{-x} dx \;\; = & \;\; \frac{\Gamma(n)}{(n\bar\delta)^n},
\label{eq:scaleint}
\end{align}
resulting in the posterior density as\footnote{Event counts $n$ are of course integers so $\Gamma(n)=(n-1)!$ but we will keep as $\Gamma(n)$ as it will cancel later.}
\begin{equation}
\pi(\lambda\mid\delta) = \frac{(n\bar\delta)^{n}}{\Gamma(n)}\;\lambda^{-(n+1)}\,e^{-n\bar\delta/\lambda}.
\label{eq:post}
\end{equation}

\sep[Marginalising the rate]
The predictive density of the next duration is obtained by averaging the exponential likelihood of $\delta'$ over the posterior of $\lambda$:
\begin{align}
p(\delta'\mid\delta)
&= \int_0^\infty p(\delta'\mid\lambda) \cdot \pi(\lambda\mid\delta)\,d\lambda = \nonumber\\
&= \frac{(n\bar\delta)^{n}}{\Gamma(n)}\int_0^\infty \lambda^{-(n+2)}\,e^{-(n\bar\delta+\delta')/\lambda}\,d\lambda.
\label{eq:predint}
\end{align}
Following a similar change of variables as in \eqref{eq:scaleint}, the integral equals $\Gamma(n+1)/(n\bar\delta+\delta')^{n+1}$, which after using the recursion $\Gamma(n+1) = n \cdot \Gamma(n)$ results in the predictive posterior density for the next duration
\begin{equation}
p(\delta'\mid\delta) = \frac{n^{n+1}\,\bar\delta^{\,n}}{(n\bar\delta+\delta')^{n+1}}.
\label{eq:pred}
\end{equation}
This is a Lomax (Pareto type II) density in $\delta'$ with shape $n$ and scale $n\bar\delta$. It has heavier tails than the exponential \eqref{eq:expdens}: marginalising over the unknown rate widens the predictive distribution, exactly as it should when $\lambda$ is estimated from a finite window. As $n\to\infty$, $\bar\delta\to\lambda$ and \eqref{eq:pred} collapses to the exponential $\lambda^{-1}e^{-\delta'/\lambda}$, recovering the known-rate case.

\subsubsection{Frequentist derivation}
\sep[Predictive likelihood with the nuisance rate eliminated]
The same density arises without a prior, as a frequentist predictive likelihood in which the nuisance parameter $\lambda$ is eliminated by maximisation rather than integration~\citep{freq1,freq2,freq3}. The maximum-likelihood estimate from \eqref{eq:like} is
\begin{equation}
\hat\lambda_{ML}(\delta) = \arg\max_\lambda \mathcal{L}(\lambda;\delta) = \bar\delta.
\end{equation}
The predictive density of $\delta'$ is built by treating $\delta'$ as one further observation and re-maximising the joint likelihood over $\lambda$:
\begin{equation}
p(\delta'\mid\delta) = \frac{p\big(\delta',\delta \mid \hat\lambda_{ML}(\delta',\delta)\big)}{\displaystyle\int_0^\infty p\big(y,\delta \mid \hat\lambda_{ML}(y,\delta)\big)\,dy},
\label{eq:freqpred}
\end{equation}
where the joint MLE over all $n+1$ points is $\hat\lambda_{ML}(\delta',\delta) = (n\bar\delta+\delta')/(n+1)$. Substituting it into the joint exponential likelihood, the numerator is
\begin{align}
p\big(\delta',\delta \mid \hat\lambda_{ML}\big)
&= \hat\lambda_{ML}^{-(n+1)}\exp\!\left(-\frac{n\bar\delta+\delta'}{\hat\lambda_{ML}}\right) = \nonumber\\
&= \left(\frac{n+1}{e}\right)^{n+1} (n\bar\delta+\delta')^{-(n+1)}.
\label{eq:freqnum}
\end{align}
The normalisation denominator integrates as,
\begin{align}
\int_0^\infty &\left(\frac{n+1}{(n\bar\delta+y)\,e}\right)^{n+1} dy = \nonumber\\
&= \left(\frac{n+1}{e}\right)^{n+1}\frac{(n\bar\delta)^{-n}}{n}.
\label{eq:freqden}
\end{align}
Dividing \eqref{eq:freqnum} by \eqref{eq:freqden}, the factors $\big((n+1)/e\big)^{n+1}$ cancel and we recover exactly
\begin{equation}
p(\delta'\mid\delta) = \frac{n^{n+1}\,\bar\delta^{\,n}}{(n\bar\delta+\delta')^{n+1}},
\end{equation}
identical to the Bayesian predictive density \eqref{eq:pred}. The agreement is reassuring, though expected. With the choices made in this modelling, the Bayesian and frequentist routes deliver the same statistic.

\subsection{The p-value of a fill}
\sep[Cumulative predictive probability]
The relevant tail probability is that of seeing a duration \emph{as short as or shorter than} the observed fill-to-print duration $\delta'$. Integrating \eqref{eq:pred},
\begin{align}
u(\delta')
\;\equiv\; P(\delta_{n+1}\le \delta'\mid\delta) =&\nonumber \\
= \int_0^{\delta'} \frac{n^{n+1}\bar\delta^{\,n}}{(n\bar\delta+\delta)^{n+1}}\,d\delta =& 1 - \left(\frac{n\bar\delta}{n\bar\delta+\delta'}\right)^{\! n}.
\label{eq:pvalue}
\end{align}
This $u(\delta')$ is the (one-sided, left-tail) $p$-value for the null hypothesis that the fill-to-print duration was generated by the same local Poisson process as the surrounding market prints. A small $u$ means the print arrived implausibly soon after the fill: strong evidence that the fill triggered it. 

Under the null, $u(\delta_{n+1})$ is uniformly distributed on $[0,1]$ -- the property we exploit next.

\section{Combining fills: Fisher's method}
\label{sec:fisher}
\sep[The need to pool evidence]
A single fill yields a single $p$-value $u$, which on its own is weak because of the large uncertainty in the estimated rate parameter from a few data points. However, we can improve the power by pooling evidence of the $k$ most recent fills into one statistic, in a way that is predictive even when no individual fill is decisive. Because the per-fill $p$-values are independent and uniform under the null, they can be combined by Fisher's method~\citep{fisher}.

\sep[Derivation of the combination law]
Let $u_1,\dots,u_k$ be the $p$-values of $k$ fills. Consider the transformation $-2\ln u_i$. Under the null each $u_i \sim \mathrm{Uniform}(0,1)$, so for $x \ge 0$
\begin{equation}
P\big(-2\ln u_i \le x\big) = P\big(u_i \ge e^{-x/2}\big) = 1 - e^{-x/2},
\end{equation}
which is precisely the CDF of a $\chi^2_2$ random variable. 

Each term $-2\ln u_i$ is therefore $\chi^2_2$, and since the fills are independent, their sum is $\chi^2$ with the degrees of freedom added:
\begin{equation}
X_k \;\equiv\; -2\sum_{i=1}^{k}\ln u_i \;\sim\; \chi^2_{2k}
\label{eq:fisher}
\end{equation}
under the null of no impact.
The combined evidence against the no-impact null is the upper-tail probability $P(\chi^2_{2k} \ge X_k)$, available in closed form for the $\chi^2$ distribution with even degrees of freedom,
\begin{equation}
P(\chi^2_{2k} \ge X_k) = e^{-X_k/2}\sum_{j=0}^{k-1}\frac{(X_k/2)^j}{j!},
\label{eq:chitail}
\end{equation}
so the running combined $p$-value can be updated incrementally as each fill arrives, without any special-function evaluation.

\sep[Why this is efficient]
Fisher's method is the natural pooling rule here because it weights each fill by the logarithm of its $p$-value: a single very surprising fill (tiny $u_i$, large $-2\ln u_i$) contributes a large amount of evidence, while many unremarkable fills accumulate slowly. The number of fills required to reach a given confidence is thus \emph{adaptive} -- governed by how informative the fills actually are. 

When liquidity is genuinely benign, $X_k$ tracks its null mean of $2k$ and never triggers. When fills are being echoed by the market, a small number of surprising durations is enough to push $X_k$ deep into the tail. In the extreme, two or three sharply surprising fills can already constitute decisive evidence, in stark contrast to the several hundred fills that slippage detection requires (Section~\ref{sec:slippage}).

\begin{figure*}[t]
\includegraphics[width=\textwidth]{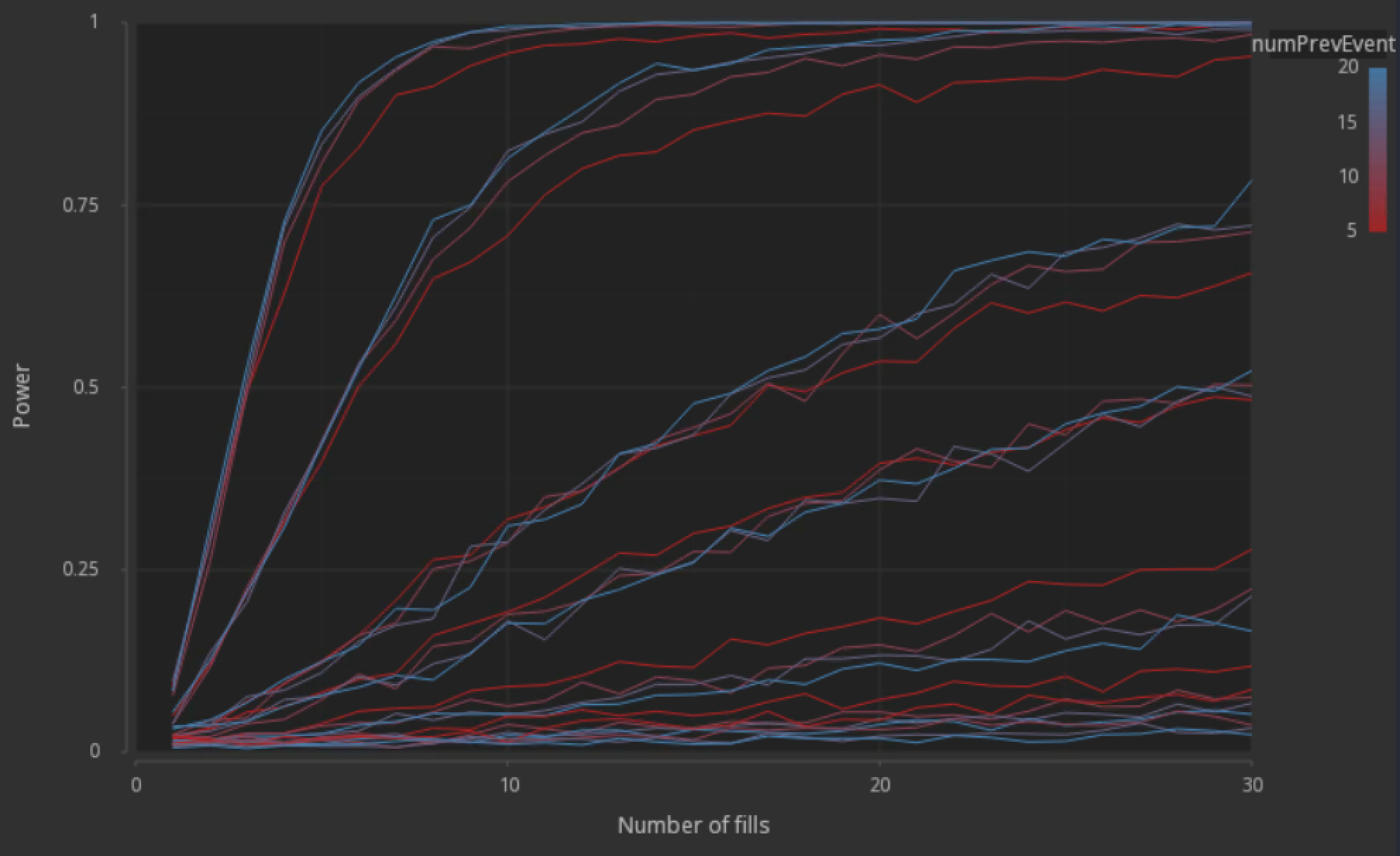}
\caption{
{\bf The method reaches significant power within a handful of fills.}
Power of the test as a function of the number of fills $k$ combined by Fisher's method \eqref{eq:fisher}. Results are from a Monte Carlo simulation in which market prints are Poisson and the durations following a fill are shortened relative to the background, so prints arrive surprisingly fast. Curve families correspond to different levels of surprise. Colour denotes the number of prior durations $n$ used to estimate the local rate ($n = 5, 10, 20$). For strong surprise, power approaches one within five to ten fills -- orders of magnitude faster than the several-hundred-fill requirement of slippage estimation (Section~\ref{sec:slippage}).
}
\label{fig:power}
\end{figure*}

\sep[Detection power]
Figure~\ref{fig:power} checks this with a simulation. We generate market prints as a Poisson process, then inject the signature impact is assumed to leave: post-fill durations shortened relative to the background, so prints arrive \emph{surprisingly fast}. We run the Fisher test \eqref{eq:fisher} at a fixed significance level and record its power as the number of fills $k$ grows. The groups of curves are different surprise levels (how much the durations are shortened) and several window sizes $n$ for the local rate estimate.

When the surprise of adverse events is large, power reaches significant levels within a few fills, in contrast to the hundreds slippage would need. A weaker surprise takes more fills. A longer window $n$ pins down the background rate more accurately and lifts power a little, at the price of adapting more slowly when that rate changes.

\section{Proposed empirical validation}
\label{sec:scope}
\sep[Two distinct claims]
It is important to be clear about what the preceding sections do and do not establish. The method rests on two logically separate claims, and they carry very different evidential burdens. 

The \emph{statistical} claim is that the post-fill duration test of Section~\ref{sec:perfill}, pooled by Fisher's method in Section~\ref{sec:fisher}, detects a departure of the post-fill timing from the locally estimated background process from only a handful of fills. This is demonstrated by Figure~\ref{fig:power}.

The \emph{economic} claim is that such a timing departure is a genuine signature of, and predictor of, adverse price impact caused by the fill. Everything operationally interesting about the method depends on the economic claim, and simulations alone cannot validate that claim.


\sep[The empirical test of the linkage]
The validation that \emph{does} bear on the economic claim correlates two quantities that are measured \emph{independently} and from \emph{different data channels}. 

The surprise statistic is computed purely from event \emph{timestamps} -- the fill-to-print durations and the surrounding background durations. A measure of realised impact -- for instance signed post-fill slippage at a short horizon, as in \eqref{eq:slippage} -- is computed purely from the \emph{price} series. Because the two are derived from disjoint inputs, a systematic relationship between them cannot be an artefact of construction. 

The primary test is therefore: across a large sample of real fills, do high-surprise fills exhibit systematically larger adverse realised price moves than low-surprise fills? A monotone relationship would be direct, non-circular evidence for the linkage; its absence would refute the method regardless of how well the statistical machinery performs.

\sep[Placebo and causal tests]
Two further tests sharpen the inference. A \emph{placebo} controls for the possibility that fast prints and price drift simply co-occur for reasons unrelated to our trading: recompute the identical surprise statistic at randomly chosen timestamps that are not our fills (or at other participants' fills, where identifiable), and verify that the surprise-to-slippage relationship is present around our own fills and absent at the placebo times. 

Finally, the only test that speaks to \emph{causation and usefulness}, rather than correlation, is a live controlled experiment: route a randomised subset of orders with the detector controlling, say trade rate, and a matched control subset without, and compare realised slippage. This is operationally costly and lies beyond the scope of a methodological exposition, but it is the decisive experiment, and the metric is designed so that such an A/B comparison is straightforward to run.


\section{Conclusion}
This paper proposes a methodology that may be useful for solving an important problem in execution. Detecting impact real-time allows the trader to adjust their actions to rapidly changing market conditions.

Perhaps more importantly, distinguishing adverse price moves caused by the trader's action from price moves due to alpha or a competing market participant allows the trader to decide whether to, in simple terms, hit the brakes or hit the accelerator.

The methodology rests on estimating the surprise factor in the timing of adverse market events following trader action. Under a locally estimated market Poisson event rate we presented the predictive-duration $p$-value, derived identically from Bayesian and frequentist arguments. Fisher's method then pools these into a single, realtime measure of evidence.

The statistic reacts within a few trader fills and is discriminating in that a few large unexpected surprises carry more evidence of impact than many small ones. By conditioning the events on adverse moves, it is pointed at impact rather than mere activity. 


The simulation establishes the statistical significance of the method. Its economic significance, on which the method's practical usefulness ultimately rests, is an empirical question about real markets which cannot be settled by a simulation.

Circling back to the full picture, every trader event produces a market reaction with two coordinates, one on the time-axis $\delta t$ (\emph{when} the next market event happens), and one on the price-axis $\delta p$ (\emph{how much} does the event impact the market). These coordinates are generated, absent impact, from a background joint probability law \[p(\delta t, \delta p) = p(\delta t) \cdot p(\delta p | \delta t).\] Impact and toxicity distort that law, and the task is to detect the distortion in real time. 

This paper has pursued one axis of it: the timing, $p(\delta t)$, where the causal question lives -- did our own action trigger the response? The other axis, the price impact $p(\delta p \mid \delta t)$, can be thought of as the familiar markout, and carries the magnitude. Neither alone is the whole story. The natural object is the \emph{joint} surprise -- a test that reads both axes at once, with a price change that is at the same time surprisingly soon \emph{and} surprisingly adverse being the sharpest evidence of caused impact. 

With adverse-event selection, the time-domain detector developed here is the first instance of that broader programme: realtime detection of impact and toxicity along the time and price axes together.

\end{document}